\documentclass[12pt,preprint]{aastex}
\usepackage[dvips]{color}
\usepackage{graphicx}

\def\msol{\hbox{\kern 0.20em $M_\odot$}}
\def\lsol{\hbox{\kern 0.20em $L_\odot$}}
\def\rsol{\hbox{\kern 0.20em $R_\odot$}}
\def\sr{\hbox{\kern 0.20em sr}}
\def\srmu{\hbox{\kern 0.20em sr$^{-1}$}}
 
\def\g{\hbox{\kern 0.20em g}}
\def\gmu{\hbox{\kern 0.20em g$^{-1}$}}
\def\kg{\hbox{\kern 0.20em kg}}
\def\pc{\hbox{\kern 0.20em pc}}
 
\def\mum{\hbox{\kern 0.20em $\mu$m}}
\def\mumd{\hbox{\kern 0.20em $\mu$m$^{-2}$}}
\def\cm{\hbox{\kern 0.20em cm}}
\def\m{\hbox{\kern 0.20em m}}
\def\km{\hbox{\kern 0.20em km}}
\def\nm{\hbox{\kern 0.20em nm}}
 
\def\s{\hbox{\kern 0.20em s}}
\def\h{\hbox{\kern 0.20em h}}
\def\sec{\hbox{\kern 0.20em sec}}
\def\min{\hbox {\kern 0.20em min}}
\def\smu{\hbox{\kern 0.20em s$^{-1}$}}
\def\smd{\hbox{\kern 0.20em s$^{-2}$}}
\def\an{\hbox{\kern 0.20em an}}
\def\anmu{\hbox{\kern 0.20em an$^{-1}$}}
\def\deg{\hbox{\kern 0.20em $^{\rm o}$}}
\def\yr{\hbox{\kern 0.20em yr}}
\def\yrmu{\hbox{\kern 0.20em yr$^{-1}$}}
\def\Myr{\hbox{\kern 0.20em Myr}}
\def\Mymu{\hbox{\kern 0.20em Myr$^{-1}$}}
\def\K{\hbox{\kern 0.20em K}}
\def\pcmu{\hbox{\kern 0.20em pc$^{-1}$}}
\def\pcmd{\hbox{\kern 0.20em pc$^{-2}$}}
\def\pcmt{\hbox{\kern 0.20em pc$^{-3}$}}
\def\kms{\hbox{\kern 0.20em km\kern 0.20em s$^{-1}$}}
\def\kmpd{\hbox{\kern 0.20em km$^{2}$}}
\def\kpc{\hbox{\kern 0.20em kpc}}
\def\cms{\hbox{\kern 0.20em cm\kern 0.20em s$^{-1}$}}
\def\erg{\hbox{\kern 0.20em erg}}
\def\ergs{\hbox{\kern 0.20em erg}}
\def\cmpd{\hbox{\kern 0.20em cm$^2$}}
\def\cmmd{\hbox{\kern 0.20em cm$^{-2}$}}
\def\cmms{\hbox{\kern 0.20em cm$^{-6}$}}
\def\cmpt{\hbox{\kern 0.20em cm$^3$}}
\def\cmmt{\hbox{\kern 0.20em cm$^{-3}$}}
\def\mpd{\hbox{\kern 0.20em m$^2$}}
\def\mmd{\hbox{\kern 0.20em m$^{-2}$}}
\def\mpt{\hbox{\kern 0.20em m$^3$}}
\def\mmt{\hbox{\kern 0.20em m$^{-3}$}}
\def\mujy{\hbox{\kern 0.20em $\mu$Jy}}
\def\mjy{\hbox{\kern 0.20em mJy}}
\def\Mj{\hbox{\kern 0.20em MJy}}
\def\jy{\hbox{\kern 0.20em Jy}}
\def\ghz{\hbox{\kern 0.20em GHz}}
\def\srmd{\hbox{\kern 0.20em sr$^{-1}$}}
\def \kms{km~$\rm{s}^{-1}$}

\def \mum{$\mu$m}

\def\G{\hbox{\kern 0.20em G}}

\def\h13cop{\hbox{H$^{13}$CO$^{+}$}}

\def\S+{\hbox{S{\small II}}}

\slugcomment{The Evolving ISM in the Milky Way \& Nearby Galaxies}

\shorttitle{Blue Compact Dwarfs in the Mid-IR}
\shortauthors{Wu et al.}

\begin{document}

\newcommand{\jfourteen}{\hbox{$J=14\rightarrow 13$}}
 \title{The Mid-Infrared Properties of Blue Compact Dwarf Galaxies}

 \author{Yanling Wu\altaffilmark{1}, V.
   Charmandaris\altaffilmark{2,3}, J.R. Houck\altaffilmark{1}, J.
   Bernard-Salas\altaffilmark{1}, V. Lebouteiller\altaffilmark{1}}

\altaffiltext{1}{Astronomy Department, Cornell University, Ithaca, NY 14853}

\altaffiltext{2}{University of Crete, Department of Physics, Heraklion, GR-71003,Greece}

\altaffiltext{3}{IESL/FORTH, Heraklion, GR-71110, Greece, \& Chercheur  Associ\'e, Obs. de Paris, F-75014, France}

\begin{abstract}
  The unprecedented sensitivity of the Spitzer Space Telescope has
  enabled us for the first time to detect a large sample of Blue
  Compact Dwarf galaxies (BCDs), which are intrinsically faint in the
  infrared. In the present paper we present a summary of our findings
  which providing essential information on the presence/absence of the
  Polycyclic Aromatic Hydrocarbon features in metal-poor environments.
  In addition, using Spitzer/IRS high-resolution spectroscopy, we
  study the elemental abundances of neon and sulfur in BCDs and
  compare with the results from optical studies. Finally, we present
  an analysis of the mid- and far-infrared to radio correlation in low
  luminosity low metallicity galaxies.

\end{abstract}


\lefthead{Wu et al.}

\righthead{Mid-Infrared Study of BCDs}

%\vspace*{-1cm}
\section{Introduction}
\vspace*{-0.2cm} Blue compact dwarf galaxies (BCDs) are dwarf galaxies
with blue optical colors resulting from one or more intense bursts of
star formation, low luminosities (M$_B>$-18), and small sizes.
Although BCDs are defined mostly by their morphological parameters,
they are globally found to have low heavy element abundances as
measured from their HII regions (1/30 -- 1/2\,Z$_\odot$, Izotov \&
Thuan, 1999). The low metallicity of BCDs is suggestive of a young
age, since their interstellar medium is chemically unevolved. However,
some low metallicity BCDs (i.e. IZw18) do display an older stellar
population and have formed a large fraction of their stars more than 1
Gyr ago (see Aloisi et al. 2007).  The plausible scenario that BCDs
are young is intriguing within the context of cold dark matter models
which predict that low-mass dwarf galaxies, originating from density
perturbations much less massive than those producing the larger
structures, can still be forming at the current epoch.  However,
despite the great success in detecting galaxies at high redshift over
the past few years, bona fide young galaxies still remain extremely
difficult to find in the local universe (Kunth \& \"{O}stlin 2000).
This is likely due to the observational bias of sampling mostly
luminous, more evolved galaxies at high redshifts.  If some BCDs are
truly young galaxies, they would provide an ideal local laboratory to
understand the galaxy formation processes in the early universe.

Accumulated observational evidence over the recent years provided more
details on the unique properties of these galaxies (for a review see
Kunth \& \"{O}stlin 2000). Early ground-based observations by Roche et
al.  (1991) on the MIR spectra of 60 galaxies revealed that Polycyclic
Aromatic Hydrocarbon (PAH) emission is generally suppressed in
low-metallicity galaxies, which could be due to hard photons
destroying the particles that produce the unidentified infrared bands.
The suppression of the PAH emission is also seen in the mid-IR spectra
of four BCDs discussed by Madden (2000), Galliano et al. (2005), and
Madden et al. (2006). More recent work based on {\em Spitzer}
observations has confirmed that PAH emission is missing in the most
metal-poor galaxies (Houck et al. 2004b; Engelbracht et al. 2005,2008;
Wu et al. 2006, 2007a; Rosenberg et al. 2006). Dwek (2005) proposed
that the delayed injection of carbon molecules into the interstellar
medium (ISM) might be partly responsible for the absence of PAH
features in young star-forming regions, or for the existence of a
metallicity threshold below which PAHs have not formed.

For our study, as part of the IRS (Houck et al. 2004a) Guaranteed Time
Observation (GTO) program, we have compiled a large sample ($\sim$60)
of BCD candidates selected from the Second Byurakan Survey (SBS),
Bootes void galaxies, and other commonly studied BCDs. We obtained
low-resolution and high-resolution 5-35\,$\mu$m spectra, as well as 16
and 22\,$\mu$m peak-up images for our sample. Full details on our
observation strategy and data reduction as well as ancillary data used
can be found in Wu et al. (2006,2007a, 2007b,2008).

%Metallicity is a key parameter that influences the formation and
%evolution of both stars and galaxies. Detailed studies of the
%elemental abundances of BCDs have already been carried out by several
%groups (Izotov \& Thuan 1999; Kniazev et al. 2003; Shi et al. 2005).
%However, because these studies were performed in the optical, they
%were limited by the fact that the properties of some of the deeply
%obscured regions in the star-forming galaxies may remain inaccessible
%due to dust extinction at these wavelengths. Infrared studies on
%elemental abundances enjoy several advantages. First, the infrared
%emission is much less affected by dust extinction problems. Second, in
%the infrared, more ionization stages of an element become available,
%thus avoids the need of ionization correction factors (ICFs). Finally,
%the infrared lines are much less sensitive to the electron temperature
%fluctuations than the corresponding optical lines of the same ion.

\vspace*{-0.5cm}
\section{Results}
\vspace*{-0.2cm}
The 5-38\,$\mu$m low-resolution spectra for 12 BCDs were published by
Wu et al. (2006) where we refer the reader for their detailed
description. In brief, the mid-IR spectra of these low-metallicity
galaxies display a variety of spectral features. PAH emission is
absent in some BCDs, such as IZw18 and SBS0335-052, while prominent PAH
emission was detected in other BCDs, including NGC1140. Fine structure
lines of [SIV]10.51\,$\mu$m, [NeII]\,12.81\,$\mu$m,
[NeIII]\,15.55\,$\mu$m and [SIII]\,18.71\,$\mu$m were detected in most
BCDs even from the low-resolution spectra and line flux measurements
for 13 BCDs from the high-resolution data can be found in Wu et al.
(2008).
 
\vspace*{-0.5cm}
\subsection{PAH emission in low metallicity galaxies}
\vspace*{-0.2cm} As mentioned earlier the absence of PAH emission
could be due to the low abundance of carbon and/or nucleating grains.
To examine a possible variation in the PAH EW with metallicity, we
plot the PAH Equivalent Widths (EWs) of 6.2 and 11.2\,$\mu$m for our
sample as a function of their metallicities. We can see that PAH
emission is absent in the most metal-poor BCDs and it appears that
there is a trend showing that galaxies with a lower metallicity may
have smaller PAH EWs (see Fig. 7 in Wu et al. 2006).

The presence of a young starburst in a low-metallicity environment
results in the production of high-energy photons that can propagate
relatively large distances before being absorbed by the metals in the
ISM. Because of the rather large difference in the ionization
potentials of Ne$^{++}$ (41 eV) and Ne$^+$ (22 eV), the ratio of
[NeIII]/[NeII] is a good tracer of the hardness of the interstellar
radiation field. We observe that the PAH EWs at 6.2 and 11.2\,$\mu$m
are generally suppressed in a harder radiation field as indicated by a
larger [NeIII]/[NeII] ratio, suggesting that the deficiency in PAH
emission may be related to the destruction of the photodissociation
region (PDR) by hard UV photons. Another important parameter of the
radiation field is its UV luminosity density. In young starbursts,
L$_{FIR}$ is representative of the total UV luminosity.  We plot the
PAH EWs as a function of its 22\,$\mu$m luminosity density and find
that there is a trend that PAH EW decreases with increasing luminosity
density. However, there is large scatter to both relations.

We have shown that there are generally weaker PAHs in metal-poor
environments. We have also discussed the effects of the hardness of
the radiation field and the luminosity density on the destruction of
PAH molecules. Is the absence of PAHs in metal-poor galaxies solely
due to formation effects, destruction effects, or some combination? To
examine this, we plot the PAH EW as a function of a new quantity:
([NeIII/[NeII])(L$_{22 \mu m}$/V)(1/Z), where the product of the neon
ratio and the luminosity density represents the destruction effect and
Z, the metallicity of the galaxy, represents the formation effect. We
find that there is a much tighter anti-correlation (Fig. 1) as
compared to other relations with PAHs that we have inspected on a
series of parameters. This suggests that the absence of PAHs in a
metal-poor environment is due to a combination of formation effects
and destruction effects. A more detailed analysis on the variation of
the various PAH bands on different environments over a large range on
metallicities can be found on (Smith et al. 2007, Engelbracht et al
2008, and Galliano et al. 2008)

\vspace*{-0.5cm}
\subsection{Elemental Abundances}
\vspace*{-0.2cm} Using the infrared fine structure lines, one can
estimate the elemental abundances for some elements, i.e. neon and
sulfur. There are several methods to derive the chemical abundances.
We use an empirical method, which derives ionic abundances directly
from the observed lines of the relavent ions. To do so, we need to
have the flux of one hydrogen recombination line, the dust extinction,
as well as the electron temperature and density of the interstellar
medium (ISM).

In the wavelength range observed by the IRS, the flux of Hu$\alpha$
(12.37\,$\mu$m) line is most commonly used (converted to H$\beta$) to
estimate the flux of ionized hydrogen. However, this line is usually
faint in BCDs and in cases where Hu$\alpha$ is not detected, we use
the radio continuum or H$\alpha$ images to derive the flux of
H$\beta$.  Since the infrared determination of elemental abundances
does not depend strongly on the electron temperatures and densities,
we adopt optically derived T$_e$ and N$_e$ from the literature. For
sources where such information is not available, we assume a typical
T$_e$ of 10,000\,K and N$_e$ of 100\,cm$^{-3}$. We also adopted the
E$_{B-V}$ values calculated from hydrogen recombination lines from
optical spectra for dust extinction, and in cases where there is
strong evidence for dust-enshrouded regions, we estimate its dust
extinction from the strength of silicate absorption features at
9.7\,$\mu$m.

In Figure 2 we plot the abundances of neon and sulfur for our BCD
sample as derived from the IRS high-resolution data. Verma et al.
(2003) have found a positive correlation between the neon and argon
abundances for their sample of starburst galaxies using ISO
observations, while their data show no correlation between the sulfur
and neon and/or argon abundances (indicated as stars on Fig. 2).
However, our sources show that the neon and sulfur abundances scale
with each other. In the same figure, we also plot the proportionality
line of the ratio for (Ne/S).  The maximum and minimum values of the
solar neon and sulfur abundances are indicated by the width of the
gray band, and we find that most of our BCDs have ratios above those
values, indicating a higher Ne/S ratio than that has been found in the
solar neighborhood.  If we compare the infrared derived Ne/S ratios
for BCDs with the optical derived Ne/S ratios, we find a discrepancy
of a factor of $\sim$2 (see Fig. 6 in Wu et al. 2008). This is
probably due to the difference in the infrared and optical methods for
estimating elemental abundances (e.g.  the uncertainties in the ICFs
adopted by optical studies).  However, the average Ne/S ratio of the
13 BCDs we studied (12.5$\pm$3.1) is consistent with the Ne/S ratios
derived in the HII regions from several other studies.

\vspace*{-0.5cm}
\subsection{The Infrared/Radio Correlation}
\vspace*{-0.2cm}

The FIR/radio correlation is known to be ubiquitous among many late
type galaxies. However, little could be done for low luminosity
systems. Hopkins et al. (2002) found that the star formation rates
(SFRs) estimated from 1.4\,GHz and 60\,$\mu$m luminosities agree with
each other for the BCDs they studied, while Wu et al.  (2005)
indicated a slope change for dwarf galaxies in a similar comparison of
SFRs. A number of deep Spitzer mid-IR and FIR surveys can now probe a
population of galaxies with low infrared luminosities, for which
ancillary data, including deep radio imaging, are becoming available.
With these new data, we will be able to investigate the connection
between the infrared and radio emission in low luminosity systems.

%With 24\,$\mu$m (22\,$\mu$m) and 70\,$\mu$m MIPS fluxes becoming
%available from {\em Spitzer}, as well as {\em IRAS} 60 and 100\,$\mu$m
%data, and radio continuum fluxes from NVSS or FIRST, we can
%investigate the infrared/radio correlation in low luminosity dwarf
%galaxies.

In Fig. 1 of Wu et al. (2007b), we plot the radio luminosity of our
sample as a function of the FIR luminosity. The luminosities of the
sources we study span nearly 4 orders of magnitudes, but the
correlation between the FIR and the radio is remarkably tight with a
scatter less that a factor of 2 and consistent with a linear fit of
with slope q$_{\rm FIR}$=1.09$\pm$0.07 which agrees well with the
slope of 1.10$\pm$0.04 found by Bell (2003) for a sample of 162
galaxies, as well as the slope of 1.11$\pm$0.02 for the infrared
selected sources from the IRAS Bright Galaxy Sample (BGS) (Condon et
al. 1991). Using the MIPS 24$\mu$m data we also examined the MIR to
radio correlation for our sample following the definition of q$_{24}$
by Appleton et al. (2004). We find that q$_{24}$=1.3$\pm$0.4, slightly
higher than the average q$_{24}$ for the galaxies in the first look
survey (Appleton et al. 2004), though still consistent within
2$\sigma$ (see Fig. 2 of Wu et al. 2007b. We search for correlations
between the q ratios of the galaxies in our sample with a few physical
parameters, such as metallicity, dust temperature, etc. We do not see
any clear correlation between the q ratios and the temperature, while
for the metallicity of our sample, we find that q$_{24}$ is generally
suppressed for metal-poor sources (12+log(O/H)$<$8.0), with
SBS0335-052E as a clear outlier (see Fig. 3). A more detailed
discussion on the q$_{24}$ ratio with metallicity can be found in Wu
et al. (2007b).

\vspace*{-0.5cm}
\section{Summary}
\vspace*{-0.2cm}
We have explored the mid-IR properties of BCDs with the {\em Spitzer}
Space Telescope. Using the low-resolution and high-resolution modules
of the IRS, we have obtained 5-35\,$\mu$m spectra for our BCD sample.
PAH emission at 6.2, 7.7, 11.2 and 12.8\,$\mu$m is detected in some
BCDs, though their strength varies substantially, and we find that the
suppression of PAHs in metal-poor harsh environment is likely due to a
combination of formation and destruction effects. Using infrared
fine-structure lines, we have also estimated the elemental abundances
of neon and sulfur for 13 BCDs. It appears Ne and S abundances are
well correlated in our sample, while their ratio is higher than the
values found in the solar neighborhood. The infrared derived Ne/S
ratio is also a factor of 2 higher than the optical derived Ne/S ratio
for the same sample, though our result is consistent with the Ne/S
found by other infrared studies in HII regions. Finally, we have also
investigated the IR/radio correlation in low luminosity systems and we
find that both the mid-IR and FIR luminosities scale with the radio
luminosities, though the scatter is larger for mid-IR/radio.

\begin{figure}
\centerline{
  \includegraphics[width=450pt,height=300pt]{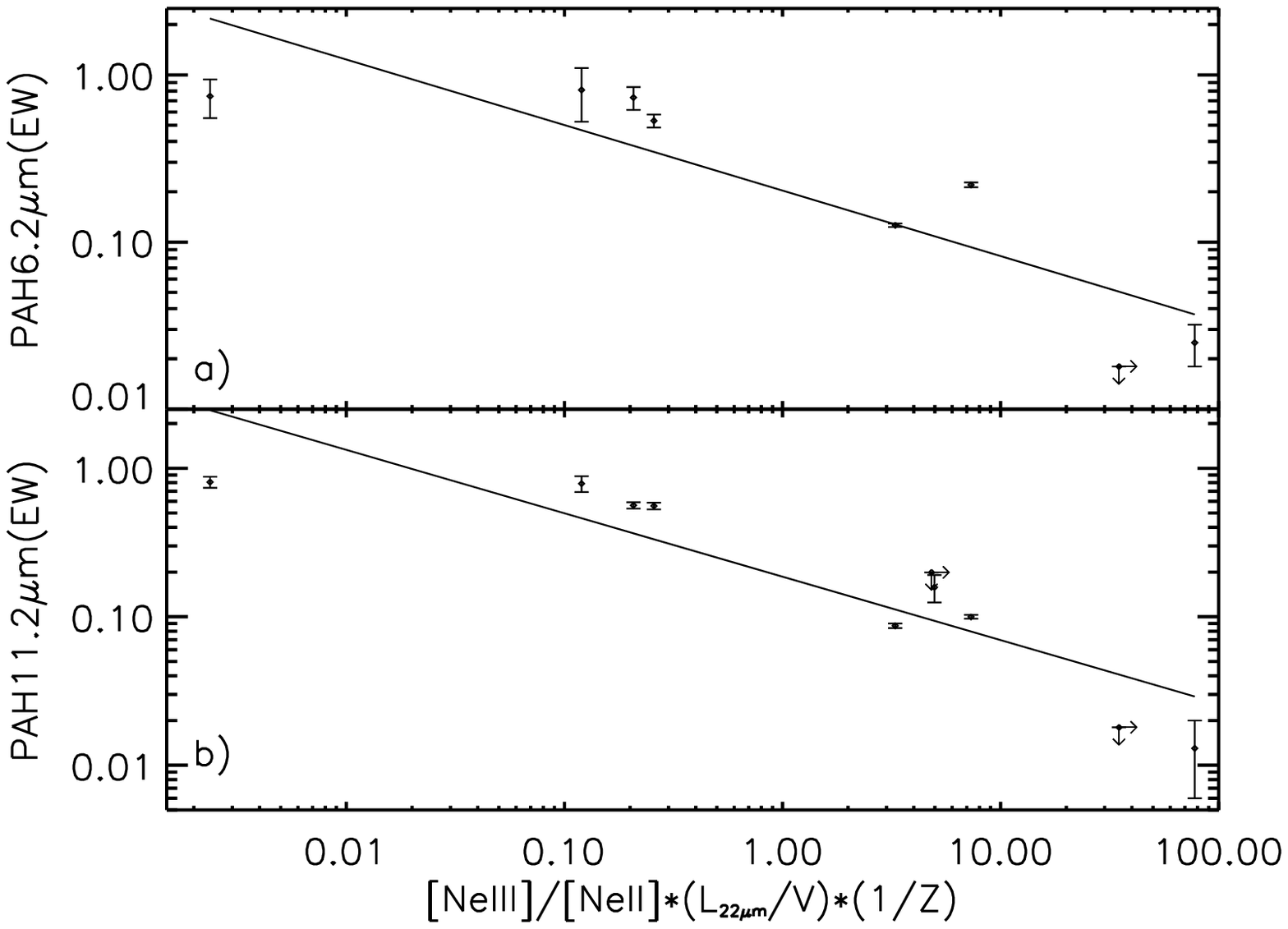}
}
\caption{\label{fig:1} a) The PAH EW at 6.2\,$\mu$m vs the product of
  the hardness of the radiation field and the luminosity density,
  divided by the metallicity of the galaxy. We can see that there is
  an anti-correlation that the PAH EW decreases when the factor
  ([NeIII]/[NeII])$\times$(L$_{22 \mu m}$/V)$\times$(1/Z) increases.
  b) Same as a) but for the 11.2\,$\mu$m PAH (see section 4.6 in Wu et
  al. 2006 for more details).}
\end{figure}

\begin{figure}
\centerline{
  \includegraphics[width=300pt,height=200pt]{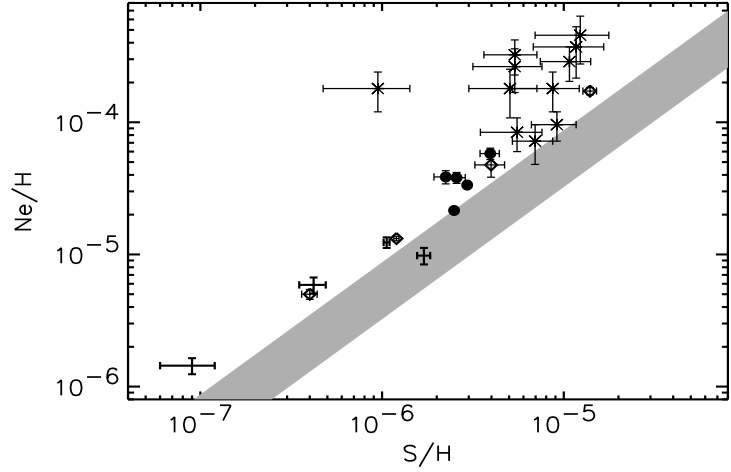}
}
\caption{\label{fig:2} A plot of the Ne/H vs S/H abundance. Our BCDs
  are indicated by the filled circles if Hu\,$\alpha$ is detected and
  by diamonds if it is not detected. The sources that are shown by
  only the error bars are those that have non-detection of [NeII] or
  [SIII]. The starburst galaxies from Verma et al. (2003) are marked
  with the stars. We use the grey band to indicate the locus on the
  plot where the ratio of Ne/S would be consistent with the different
  solar values for neon and sulfur abundances (see section 4.2 in Wu
  et al. 2008 for more details).}
\end{figure}

\begin{figure}
\centerline{
  \includegraphics[width=300pt,height=200pt]{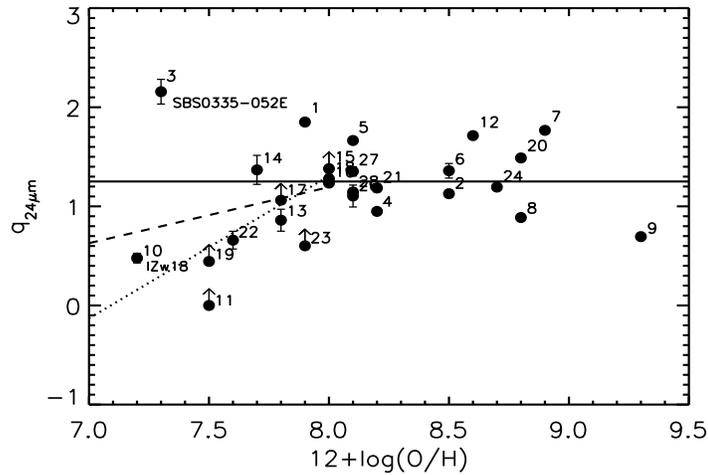}
}
\caption{\label{fig:3} The q$_{24}$ ratio plotted as a function of the
  oxygen abundance of the BCDs. The mean q$_{24}$ value for all the
  sources is indicated by the solid line. A fit to the low metallicity
  sources (12+log(O/H)$\le$8.0) is indicated by the dashed line; while
  the dotted line is the same fit excluding SBS0335-052E (see section
  3.2 in Wu et al. (2007b) for more details).}
\end{figure}

\end{document}